\documentclass[12pt,letterpaper]{article}

\def\be{\begin{equation}}
\def\ee{\end{equation}}

\def\lbldef#1#2{\expandafter\gdef\csname #1\endcsname {#2}}

\begin{document}
\baselineskip=16pt
\pagestyle{plain}
\setcounter{page}{1}


\begin{titlepage}

\begin{flushright}
hep-th/0309041 \\
CERN-TH/2003-196\\
TAUP-2745-03 
\end{flushright}
\vfil

\begin{center}
{\Large\bf Gravitational F-terms of ${\cal N}=1$ Supersymmetric
$SU(N)$ Gauge Theories}
\end{center}

\vfil
\begin{center}

{\sc Harald Ita}$^{a\,b}$
\footnote{e-mail: {\tt ita@post.tau.ac.il, Harald.Ita@cern.ch}}, 
{\sc Harald Nieder}$^{a\,b}$
\footnote{e-mail: {\tt harald@post.tau.ac.il, Harald.Nieder@cern.ch}} 
and {\sc Yaron Oz}$^{a\,b}$ 
\footnote{e-mail: {\tt yaronoz@post.tau.ac.il, Yaron.Oz@cern.ch}}\\

\vspace*{1cm}

{\it  {$^{a}$ Raymond and Beverly Sackler Faculty of Exact Sciences\\
School of Physics and Astronomy\\
Tel-Aviv University , Ramat-Aviv 69978, Israel}} \\
\vspace*{5mm}
{\it {$^{b}$Theory Division, CERN \\
CH-1211 Geneva  23, Switzerland}}\\

\end{center}

\vfil

\begin{center}
{\large Abstract}
\end{center}

We use the generalized Konishi anomaly equations and R-symmetry anomaly
to compute
the exact perturbative and non-perturbative gravitational F-terms of
four-dimensional ${\cal N}=1$ supersymmetric gauge theories.
We formulate the general procedure for computation and 
consider chiral and non-chiral $SU(N)$ gauge theories.

\vfil
\begin{flushleft}
September, 2003
\end{flushleft}
\end{titlepage}
\newpage

\renewcommand{\baselinestretch}{1.2}  


\section{Introduction}

An exact computation of  gravitational F-terms of ${\cal N}=1$ supersymmetric gauge
theories
coupled to ${\cal N}=1$ supergravity in four dimensions is of much theoretical
importance.
Recently, 
Dijkgraaf and Vafa suggested a matrix model description where the gravitational
F-terms can be computed by summing up the 
non-planar matrix 
diagrams \cite{Dijkgraaf:2002dh}.
The assumption made is that the relevant fields are the glueball superfields $S_i$ and the
F-terms are holomorphic couplings of the glueball superfields to gravity.
The DV matrix proposal has been proven diagrammatically
in \cite{Ooguri:2003qp,Ooguri:2003tt}.

In this paper we consider gravitational F-terms of the form
\begin{eqnarray}
\Gamma_G & = & 
\int d^4x d^2\theta G_{\alpha \beta \gamma}
G^{\alpha \beta \gamma} 
 F_1 (S_i) \ ,
\label{Gammag2}
\end{eqnarray}
where $G_{\alpha\beta\gamma}$ is the  ${\cal N}=1$ 
Weyl superfield.
According to the DV proposal, $F_1 (S_i)$ is the partition function 
of the corresponding matrix model evaluated by
summing the genus one diagrams with $S_i$ being the 't Hooft
parameter.
In a non-trivial graviphoton background these F-terms
belong to a series of corrections given by
\begin{eqnarray}
\label{Gammag3}
\hat{\Gamma}_G & = & \sum_{g=1}^{\infty} g \int d^4x d^2\theta G_{\alpha \beta \gamma}
G^{\alpha \beta \gamma} (F_{\delta \zeta}F^{\delta \zeta})^{g-1}
 F_g (S_i) \ , 
\end{eqnarray}
where now $F_{\delta \zeta}$ is the graviphoton field strength
and the $F_g (S_i)$ are the genus $g$ contributions to the matrix model partition function.
When the graviphoton field strength
vanishes,
only 
the $g=1$ term in (\ref{Gammag3}) remains. 

In this paper
we will compute such gravitational F-terms in a 
trivial graviphoton background
for chiral and non-chiral supersymmetric gauge theories.
The matrix model approach to gravitational 
F-terms has been recently studied in \cite{Klemm:2002pa,Dijkgraaf:2002yn}.
The approach we will take is to use  the generalized
Konishi anomaly equations.
Such a technique has been applied in \cite{David:2003ke} 
to gauge theory with matter
in the adjoint representation of $U(N)$ and extended to the
case with a non-trivial graviphoton background in \cite{Alday:2003ms}. 
The gauge groups $SO/Sp$ with various matter contents were considered in \cite{Alday:2003dk}.
In the models that we will consider the matter fields transform in the
fundamental and antisymmetric
representations of the gauge groups,
 and we will be able to perform an exact computation.

We will formulate the general procedure for computation and 
consider chiral and non-chiral $SU(N_c)$ gauge theories.
In the analysis we will need to consider the perturbative and non-perturbative
F-terms in the glueball superfields.
The perturbative series
in the glueball superfields
will be  computed by solving a set of differential equations with respect
to the tree level couplings and using the generalized
Konishi anomaly equations. 
An important  ingredient in the analysis is 
the gravitationally
deformed chiral ring of the gauge theory.
The computation of the non-perturbative terms 
will require the use of the $U(1)_R$ anomaly.

The paper is organized as follows.
In section 2 we discuss the classical chiral ring of ${\cal N}=1$ supersymmetric gauge
theories and its deformation when coupled to  ${\cal N}=1$ supergravity.
We then consider the classical and quantum generalized Konishi relations in
the supergravity background. 
In section 3 we present the procedure for the computation of the F-terms.
The analysis has two parts. One part is the computation of the perturbative series
in the glueball superfield.
This can be done by solving a set of differential equations with respect
to the tree level couplings while using the generalized
Konishi anomaly equations. 
The second part is the computation of the non-perturbative terms in the glueball superfield.
Here we will use the requirement that the $U(1)_R$ anomaly has to be reproduced
correctly.
In section 4 we consider 
${\cal N}=1$ super Yang-Mills with non-chiral matter  in the
fundamental representation
 and a tree level
superpotential. We will compute the exact F-terms both in un-Higgsed and Higgsed branches
of the theory.
We will then consider the relation of the computation
to the vector model \cite{Argurio:2002xv}.
In section 5 we will perform a similar analysis with chiral matter
 in the fundamental and antisymmetric representations.
In appendices A and B  we provide details of computations for the non-chiral and chiral
matter models, respectively.

\section{The Deformed Chiral Ring}

In this section we will discuss the classical chiral ring of ${\cal N}=1$ supersymmetric gauge
theories and its deformation when coupled to  ${\cal N}=1$ supergravity.
We will then consider the classical and quantum generalized Konishi relations in
the supergravity background. 
\subsection{Classical Chiral Ring}

Consider first an  ${\cal N}=1$ supersymmetric 
gauge theory in flat space with a gauge group $G$ and some matter supermultiplets.
We will denote the four-dimensional Weyl spinor supersymmetry generators by 
$Q_{\alpha}$ and $\bar{Q}_{\dot{\alpha}}$.
Chiral operators are operators annihilated by
$\bar{Q}_{\dot{\alpha}}$.
For instance, the lowest component $\phi$ of a chiral superfield
$\Phi$ is a chiral operator. 
The OPE of two chiral operators is nonsingular and allows for the definition 
of the product of two chiral operators. The product of chiral operators is 
also a chiral operator. 
Furthermore, one can define a ring structure on the set of 
equivalence classes of chiral operators modulo operators of the form
$\{\bar{Q}_{\dot{\alpha}}$,$\cdots \left.\right]$.

Denote by $V$ the vector superfield in the adjoint representation
of $G$, by $\Phi$ chiral superfields in a representation 
$r$ of $G$ and by $\phi$ their lowest component. 
The field strength (spinor) superfield is $W_\alpha =
-\frac{1}{4}\bar D^2e^{-V}D_\alpha e^V$ and  is a chiral superfield.
One has
\begin{eqnarray}
\{W_\alpha^{(r)},W_\beta^{(r)}\}= 0\, , \quad W_\alpha^{(r)}\phi^{(r)} = 0
\ ,
\label{rel0} 
\end{eqnarray}
modulo $\{\bar{Q}_{\dot{\alpha}}$,$\cdots \left.\right]$ terms, 
where we noted that $\phi$ transforms in a 
representation $r$ of the
gauge group $G$ such that $W_\alpha^{(r)}= W_{\alpha}^aT^a(r)$ with $T^a(r)$
being the generators of the gauge group $G$ in the representation $r$.

Consider next the coupling of the supersymmetric gauge theory to 
a background ${\cal N}=1$
supergravity.
We denote by $G^{\alpha\beta\gamma}$ the  ${\cal N}=1$ 
Weyl superfield.
In the following we will denote by $W_\alpha$ 
the supersymmetric field strength as well as its lowest component, 
the gaugino, and similarly for $G^{\alpha\beta\gamma}$.
The chiral ring relations (\ref{rel0}) are deformed to
\begin{eqnarray}
\{W_\alpha^{(r)},W_\beta^{(r)}\}=2G_{\alpha\beta\gamma}W^{\gamma{(r)}}\, , 
\quad W_\alpha^{(r)}\phi^{(r)}=0 \ .
\label{rel1} 
\end{eqnarray}
Together with  Bianchi identities of $N=1$ supergravity these relations generate 
all the relations in the deformed chiral ring.
Some 
relations that will be used later are \cite{David:2003ke}
\begin{eqnarray}
\label{ringrelgrav}
 && \left[W^2,W_\alpha \right]=0,\quad W^2W_\alpha
=-\frac{1}{3}G^2 W_\alpha,\quad W^2W^2=-\frac{1}{3}G^2 W^2 \ ,\\
 && G^4=(G^2)^2=0 \ . \nonumber
\end{eqnarray}
Throughout the paper we will follow the conventions used in \cite{David:2003ke}.

In addition to the above kinematical relations one has
kinematical relations for the matter fields and dynamical relations 
from the variation of the tree level superpotential $W_{tree}$
\begin{eqnarray}
  \phi\frac{\partial W_{tree}}{\partial \phi}=0 \ .
\label{chirrelc}
\end{eqnarray}

\subsection{Konishi Anomaly Relations}

The classical chiral ring relations are, in general, modified
quantum mechanically.
The
classical relations arising 
from (\ref{chirrelc}) have a natural 
generalization as anomalous Ward identities of the quantized matter 
sector in a classical gauge(ino) and supergravity background.
The classical Konishi equation reads
\begin{eqnarray}
\bar D^2J=\phi'\frac{\partial W_{tree}}{\partial \phi} \ ,
\end{eqnarray}
where $J$ is the generalized Konishi current and $\delta \phi =
\phi'(\phi)$
is the  generalized Konishi transformation.
This relation gets an anomalous contribution in the quantum
theory. It
takes the form 
\cite{Cachazo:2002ry,Konishi:1983hf,Konishi:1985tu}
\begin{eqnarray}
  \label{genkonishi}
\bar D^2J=
\phi'_i\frac{\partial W_{tree}}{\partial \phi_i}+\frac{1}{32\pi^2}\left(W_\alpha{}_i{}^j
W^\alpha{}_j{}^k+\frac{1}{3}G^2\delta^k_i\right)\frac{\partial\phi'_k}{\partial\phi_i}
\ ,
\end{eqnarray}
where $i$, $j$ and $k$ are gauge indices and their contraction
is in the appropriate representation. 

Since the divergence $\bar D^2J$ is $\bar{Q}$-exact it vanishes in a 
supersymmetric vacuum. Taking the expectation value of 
(\ref{genkonishi}) in a slowly varying gaugino background $S$  we get 
the Konishi relations in a supergravity 
background given by $G^2$ 
\begin{eqnarray}
 \label{chirreld}
\left< \phi'_i\frac{\partial W_{tree}}{\partial \phi}\right>_S
+\left<\left(\frac{1}{32\pi^2}W_\alpha{}_i{}^j
W^\alpha{}_j{}^k+\frac{1}{32 \pi^2}\frac{G^2}{3}
\delta^k_i\right)\frac{\partial\phi'_k}{\partial\phi_i}\right>_S
= 0 \ . 
\end{eqnarray}
We will use this relation to determine the supergravity 
corrections to the chiral correlators, which in turn can be 
integrated to give the perturbative part of the gravitational
F-terms of the corresponding ${\cal N}=1$ gauge theory.

In the next section we will discuss the gravitational F-terms
in more detail and we will outline the methods used to calculate the
perturbative part of the gravitational couplings. We will also 
propose a non-perturbative completion
of the gravitational couplings  analogous to
the non-perturbative completion of the effective superpotential
by the Veneziano-Yankielowicz potential \cite{Veneziano:1982ah}. 
The proposal will be based on $U(1)_R$ anomaly considerations.

\section{Gravitational F-terms of ${\cal N}=1$ Gauge Theories}

We are interested in the low energy description 
of a four dimensional  ${\cal N}=1$ supersymmetric gauge theory in the
background of  ${\cal N}=1$ supergravity. 
The assumption is that the relevant field is the glueball superfield $S$ and the
F-terms are holomorphic couplings of the glueball superfield to gravity.

In the absence of supergravity, the only relevant F-term is the effective glueball
superpotential 
\begin{equation}
\Gamma_0=\int d^4x d^2 \theta W_{eff}(S) \ ,
\end{equation}
where 
\begin{equation}
S=-\frac{1}{32 \pi^2}Tr\, W_{\alpha}W^{\alpha} \ .
\end{equation}
In the matrix model description $\Gamma_0$ is computed by summing up planar diagrams and 
adding a non-perturbative   Veneziano-Yankielowicz superpotential.

When coupled to supergravity
there is a  gravitational F-term
of the form 
\begin{eqnarray}
 \label{eq:GC1} 
  \Gamma_1=\int d^4x d^2 \theta W_1(S) G^2 \ .
\end{eqnarray}
In the matrix model description it is computed by summing up non-planar diagrams and 
adding a non-perturbative contribution. 
Note that terms with higher powers of $G$ vanish due to the chiral ring relation 
(\ref{ringrelgrav}).

Our aim in this paper will be to determine the $S$ dependence of the gravitational couplings
(\ref{eq:GC1}).
We will consider chiral and non-chiral $SU(N_c)$ gauge theories with matter in the fundamental
and antisymmetric
representations of the gauge group.

\subsection*{Computation of $W_1(S)$}
\label{pertpartcorr}

Consider the supersymmetric gauge theory with  a tree level superpotential
\begin{eqnarray}
\label{treelevel}
W_{tree}=\sum_I g_I \sigma_I \ ,
\end{eqnarray}
where  $\sigma_I$ are gauge invariant chiral operators and $g_I$ the
tree level couplings.
The gradient equations for the 
holomorphic part of the effective action read
\begin{eqnarray}
  \label{gradeqn}
\frac{\partial \left(W_{eff}+G^2 W_1 \right)}{\partial g_I}=\left< \sigma_I\right>_S  \ .
\end{eqnarray}
The expectation values are taken in a slowly varying (classical)
gaugino and gravitino background.

As first discussed in \cite{Cachazo:2002ry}, for a gauge theory in 
the absence of a supergravity background the Konishi relations (\ref{chirreld})
can be used to solve for the expectation values $\left< \sigma_I\right>_S$
as a function of $S$ and the tree level couplings. One can then integrate
(\ref{gradeqn}) to determine the dependence of $W_{eff}$ on the tree level couplings.
For the gravitational coupling $W_1(S)$ a similar reasoning applies. However,
we will have to take into account the effects of the
supergravity background on the correlators of chiral operators \cite{David:2003ke}.

\bigskip\noindent{\it The Perturbative Part of $W_1(S)$}
\smallskip

In the absence of gravity the correlators of chiral operators
factorize
\begin{eqnarray}
  \label{eq:mesoncorr}
  \left< \sigma_I\sigma_J \right>&=&\left< \sigma_I \right>\left<
  \sigma_J\right> \ .
\end{eqnarray}
In the matrix model description this is the feature of the planar limit.
Here and in some of the equations in the following 
we omit for simplicity the subscript $S$.
Eq. (\ref{eq:mesoncorr}) can be used in the relations (\ref{chirreld}) in order to solve 
for $\left< \sigma_I\right>_S$ as a function of $(S,g_I)$.

However, in the presence of supergravity 
the chiral correlators do not factorize (non-planarity in the matrix model 
description) and instead we have 
\begin{eqnarray}
  \label{eq:mesoncorr2}
  \left< \sigma_I\sigma_J \right>&=&\left< \sigma_I\sigma_J \right>_c
                      +\left< \sigma_I \right>\left< \sigma_J\right>
                      \ ,  
\end{eqnarray}
with analogous relations for correlators with more chiral operators.
Also, the one point functions have to be expanded in $G^2$ as  
\begin{eqnarray}
  \label{eq:epans}
  \left< \sigma_I\right>&=& \left<\sigma_I\right>_1+ G^2\left<
  \sigma_I\right>_2 \ .
\end{eqnarray}
Note that this expansion is exact in the chiral ring due to the fact that 
$G^4$ vanishes modulo ${\bar D}$ exact terms. 
Thus, we have to express $\left< \sigma_I\right>_1$ and  $\left< \sigma_I\right>_2$ 
as functions of $S$ and $g_I$.

In the next section we will show explicitly that there are enough  
relations (\ref{chirreld}) to solve for $\left< \sigma_I
\right>,\left< \sigma_J\right>$ as well as 
for the connected correlators $\left< \sigma_I\sigma_J \right>_c$.
The perturbative part of the gravitational coupling $W_1(S)$ is then obtained by
integrating the gravitational contribution $\left< \sigma_K\right>_2$
in (\ref{eq:epans})
for the $\left< \sigma_K\right>$ appearing in the tree level potential
(\ref{treelevel}), with respect to the couplings $g_K$.

A crucial  ingredient in the analysis is the fact that 
connected correlators of three or more chiral operators vanish in the 
gravitationally deformed chiral ring. 
Let us show this property diagrammatically
in the case of massive matter. For the
chiral model the diagrammatic argument is less straightforward and 
we will invoke a different one.
 
\bigskip\noindent{\it Connected Correlators vs. $G^2$}
\smallskip

A generic n-point function can first be expanded via
its connected parts, due to its Feynman-graph interpretation
\begin{eqnarray}
  \label{eq:feynsplit}
  \left< \sigma_I\sigma_J\sigma_K...\right>=
\left< \sigma_I\sigma_J\sigma_K...\right>_c+
\left< \sigma_I\right>\left< \sigma_J\sigma_K...\right>_c+...+
\left< \sigma_I\right>\left< \sigma_J\right>\left< \sigma_K\right>...
\end{eqnarray}
We will show that the connected n-point function can be estimated by
\begin{eqnarray}
  \label{eq:conne}
  \left< \sigma_{I_1}\sigma_{I_2}...\sigma_{I_n}\right>_c\sim G^{2(n-1)} \ .
\end{eqnarray}
Recall that already $G^4$ vanishes in the chiral ring.

The proof is done perturbatively in the couplings $g_K$. 
First note that the order in $G^2$ goes up with the genus of the diagram, 
such that the lowest power in $G^2$ comes from planar diagrams. 
Second, all planar diagrams can be generated from a planar one loop diagram,
 by inserting vertices on its lines and connecting them (in a planar way). 
Operations like this do not change the order of $G^2$ of a diagram. 
Thus to estimate the order of $G^2$ of a given correlator it suffices
 to estimate the simplest one loop diagram. 

Note that in this way the interaction terms of the theory do not enter the estimate.
 The power of $G^2$ is determined by the genus of a diagram and the number of insertions in a correlator.

We consider in the following non-chiral matter in the fundamental
representation,
$(\tilde Q_i^a,Q_a^i)$.
$a=1,...,N_c$ are color indices 
and $i,j=1,...,N_f$ the flavor 
indices.  
The meson is defined by $M_i^j=\tilde Q_i^aQ_a^j$.
Matter in the adjoint has been analysed in \cite{David:2003ke}.
We estimate the lowest order contribution in the couplings expansion. 
The diagram is a circle with $n$ legs running to the $n$ insertions $M(x_i)$. 
At the leg of $M(x_i)$ the momentum $p_i$ is leaving the circle. 
We have
\begin{eqnarray}
& & \left< M(x_1)M(x_2)...M(x_n)\right>_c\sim \nonumber \\
& &(n-1)!\int\prod_{i=1}^n (\frac{d^4p_id^2\pi_i}{(2\pi)^4}e^{-ip_ix_i})
 \times\delta^4(\sum_{i=1}^n p_i)\delta^2(\sum_{i=1}^n \pi_i) 
\int \prod_{i=1}^n ds_i \int d^4kd^2\kappa \times  \\
 & &\times tr\left(e^{\left[-s_1((k-p_1)^2+W(\kappa-\pi_1)+m)-s_2((k-p_1-p_2)^2+W(\kappa-\pi_1-\pi_2+m)
-...-s_n(k^2+W\kappa+m) \right]}\right) \ , \nonumber
\end{eqnarray}
where we used the a Schwinger form of the matter propagator. 
Integrating over $p_i$ and $k$, while putting $x_i=x_j$, we remain with the fermionic integrations 
\begin{eqnarray}
 & &\frac{1}{(2\pi)^{2n}}\frac{1}{2^{2n}}\int \prod_{i=1}^n (\frac{ds_ie^{-s_im}}{s_i^2}) 
\int\prod_{i=2}^n (d^2\pi_i)d^2\kappa \\ 
& &tr\left[\left(1-s_1W(\kappa+\pi_2+...\pi_n)+
(s_1W(\kappa+\pi_2+...\pi_n))^2/2\right)...
\left(1-s_nW\kappa+(s_nW\kappa)^2/2\right) \right] \ . \nonumber
\end{eqnarray}
The fermionic integrals lead to
a cancelation of the parameters $s_i$ terms in the denominator and we get
\begin{eqnarray}
   \frac{1}{(2\pi)^{2n}}\frac{1}{2^{2n}}\left(\frac{W^{2}}{2}\right)^n\int \prod_{i=1}^n 
(\frac{ds_ie^{-s_im}}{s_i^2}) \prod_{i=1}^n (s_i^2)&=&\frac{1}{(2\pi)^{2n}}
\frac{1}{2^{3n}}\left(\frac{W^{2}}{m}\right)^n.
\end{eqnarray}
Using the ring relations then gives
\begin{eqnarray}
  \left< M(0)^n\right>_c\sim\frac{(n-1)!}{(2\pi)^{2n}}\frac{1}{2^{3n}3^{n-1}m^n}tr(W^2)G^{2(n-1)} \ .
\end{eqnarray}

\bigskip\noindent{\it Nonperturbative Part of $W_1(S)$ via Anomaly}
\smallskip

The procedure outlined above determines $W_1(S)$ up to an integration constant 
independent of the couplings $g_I$. We will argue below that 
this integration constant can be determined by the one loop
exact $U(1)_R$ anomaly.
We will fix the integration constant for generic 
effective superpotentials by matching to the one calculated below, 
in the limit that all couplings except for the mass are sent to zero.

Consider an $SU(N_c)$ ${\cal N}=1$ supersymmetric gauge theory with  $n_{F}$
chiral multiplets in the fundamental representation.
In our notations,  $SU(N_c)$ super Yang-Mills with
$N_f$ flavors has $n_{F}=2N_f$.
The index of the fundamental representation is $C(fund)=1$  
\footnote{Here we find it convenient not to distinguish between 
fundamental and antifundamental representations.} . 

In our conventions 
the anomaly ${\cal A}$ of the R-symmetry is given by 
(see e.g. \cite{McArthur:1983fk,Buchbinder:im,
Anselmi:1997am})
\begin{eqnarray}
  \label{eq:suptrace}
 {\cal A}&=&-\frac{1}{3}G^2\left[N_c^2-1-\frac{n_{F}N_c}{3}\right]+S\left[2N_c-\frac{n_{F}}{3}\right].
\end{eqnarray}
Note that here and in the following whenever we write $G^2$ we will assume it
to be normalized appropriately $G^2 \to G^2/32 \pi^2$. 
Also, contributions to the anomaly from gravitinos in the loop are not included, 
since we treat gravity as a background.
It is easy to see from (\ref{eq:suptrace}) that the gauginos have R-charge equal to one
and the chiral multiplets have R-charge equal to $\frac{2}{3}$ such that
their fermionic components have R-charge equal to $-\frac{1}{3}$.

The effective potential including the gravitational coupling that reproduces the
anomaly (\ref{eq:suptrace}) is given by
\begin{eqnarray}
  \label{eq:vycorr}
  \Gamma(S,G^2)&=&N_cS\left[-\log\frac{S}{\Lambda^3}+1\right]-\frac{n_{F}}{2}S\log\frac{\Lambda}{m}+\\
  &&+\frac{1}{6}G^2 \left[(N_c^2-1)\log\frac{S}{\Lambda_0^3}+ n_F N_c\log\frac{\Lambda_0}{m}\right] \ .
\nonumber
\end{eqnarray}
Explicitly, the transformations of the potential (\ref{eq:vycorr}) under the $U(1)_R$ with
\begin{equation}
S(x,\theta,\bar\theta)\rightarrow e^{-2i\alpha}S(x,e^{i\alpha}\theta,e^{-i\alpha}\bar\theta) \ ,
\end{equation} 
and 
\begin{equation}
G^2(x,\theta,\bar\theta)\rightarrow e^{-2i\alpha}G^2(x,e^{i\alpha}\theta,e^{-i\alpha}\bar\theta) \ ,
\end{equation} 
reproduce the Yang-Mills fields contribution to the anomaly (\ref{eq:suptrace})
\begin{eqnarray}
  \label{}
\delta_{U(1)_R} \Gamma(S,G^2)=-\frac{1}{3}(N_c^2-1)G^2+2N_cS \ .
\end{eqnarray}
The dependence on the scales of the effective potential cannot be 
fixed by the anomalous transformation properties. 
The generalized Konishi anomaly relations  will constrain the 
scale dependence. 
In (\ref{eq:vycorr}) there are four places  
where the scales $\Lambda$ or $\Lambda_0$ can appear. 
The generalized Konishi anomaly requires that the scales of each term at a given order 
in the gravitational expansion in $G^2$ have to be the same. 
We will discuss this in more detail in section \ref{branchesSQCD}.

\section{$SU(N_c)$ with Fundamental Matter}

In this section will consider $SU(N_c)$ ${\cal N}=1$ supersymmetric gauge
theory with matter in the fundamental representation $(\tilde Q_i^a,Q_a^i)$
and with a tree level superpotential
\begin{eqnarray}
  \label{eq:treesuppot}
  W_{tree}(M)=mtr(M)+\lambda tr(M^2) \ ,
\end{eqnarray}
with the meson $M_i^j=\tilde Q_i^aQ_a^j$. Here $a=1,...,N_c$ are color indices 
and $i,j=1,...,N_f$ the flavor 
indices.  

For this model we will calculate the  
the gravitational F-term. We will consider first the case  $N_f=1$ and then the general  $N_f$. 
We will then compare the analysis to the vector model calculation. 

\subsection{Generalized Konishi Equations}

Consider first the case $N_f=1$.
The theory with a tree level superpotential 
given by (\ref{eq:treesuppot}) has two classical vacua at $M=0$ and $M=-\frac{m}{2\lambda}$. 
Classically, in the first vacuum the gauge group is unbroken, 
in the second it is Higgsed to $SU(N_c-1)$.

The Konishi variations $\delta Q_i=Q_i$ and $\delta Q_i=Q_iM$  lead to the equations
\begin{eqnarray}
\label{KonishirelSQCD}
 S&=&m \left< M\right>+2\lambda \left< M^2\right>+\frac{N_c}{3}G^2,\\
 S\left< M\right> &=&m \left< M^2\right>+2\lambda \left< M^3\right>+\frac{N_c+1}{3}\left< M\right> G^2,
\nonumber
\end{eqnarray}
respectively. We note that $S$ and $G^2$ are treated as background fields such that 
we need not worry about connected correlators including either $S$ or $G^2$. 
If the correlators $\left< M^n\right>$ factorized, 
the above relations would force $G^2$ to vanish, as expected. 
In a supergravity background we have to expand 
the correlators in powers of $G^2$, 
as discussed in section \ref{pertpartcorr} 
and use the ring relations (\ref{ringrelgrav}) to reduce the number of unknowns. 
Let us do that explicitly.

The expansions of the correlators of $M$ are
\begin{eqnarray}
\label{psfactor}
\left< M M\right>&=&\left< MM\right>_c+\left< M\right>\left< M\right>,\\
\left< MMM\right>&=&3\left< MM\right>_c\left< M \right>+\left< M\right>^3, \nonumber
\end{eqnarray}
with
\begin{eqnarray}
\label{psfactor2}
\left<  M\right> & = & M_0+M_1G^2 \ , \\
\left<  M^2\right>_c\sim G^2 \ ,  & &  \left<  M^3\right>_c\sim G^4 \ . \nonumber 
\end{eqnarray}
The unknowns 
are $M_0,M_1$ and $\left<  M^2\right>_c$.
Inserting  the above expansions in (\ref{KonishirelSQCD})
leads to equations at order $O(G^0)$ and
order $O(G^2)$. They are solved by
\begin{eqnarray}
  \label{eq:sol}
M_0&=&\frac{-m \pm {\sqrt{m^2 + 8\,S\,\lambda }}}{4\,\lambda }, \nonumber \\
M_1&=&\frac{-m \pm \left( 1 - 2\,N_c \right) \,
{\sqrt{m^2 + 8\,S\,\lambda }}}{6\,\left( m^2 + 8\,S\,\lambda  \right) },\\
\left<  M^2\right>_c&=&\frac{m^2 + 
8\,S\,\lambda  \mp m\,{\sqrt{m^2 + 8\,S\,\lambda }}}{-12\,m^2\,\lambda  - 96\,S\,{\lambda }^2} \ .
\nonumber
\end{eqnarray}
The two branches
of the square root in (\ref{eq:sol}) correspond to the two classical vacua, 
with $M=0$ and $M=-\frac{m}{2\lambda}$. This can be seen by using  the $S\rightarrow0$ limit.
Note that the fact that the connected part of the cubic correlator
$\left<M^3 \right>_c$ vanishes in the chiral ring has been used by us
in solving (\ref{KonishirelSQCD}). 

Considering the more general variations
$\delta Q_i =Q_i M^k$ one can derive the relations
\begin{equation}
S\left< M^k\right> = m \left< M^{k+1}\right>+2\lambda \left< M^{k+2}\right>
+\frac{N_c+k}{3}\left< M^k\right> G^2 \ . 
\end{equation}
It is straightforward to check that these equations
are solved by $M_0,M_1$ and $\left<M^2 \right>_c$ given by
(\ref{eq:sol}) with $\left<M^n\right>_c =0$ for $n \ge 3$.

\subsection{Gravitational F-term}

Using (\ref{gradeqn})
and the form of the tree level superpotential (\ref{eq:treesuppot}) we
get for the effective superpotential $W_{eff}$ 
and the gravitational F-term $W_1$ the differential equations
\begin{eqnarray}
  \label{eq:diffeffpot}
  \frac{\partial W_{eff}}{\partial m}=\left. \left<  M\right>\right|_{O(G^0)},&& \quad 
  \frac{\partial W_{eff}}{\partial \lambda}=\left. \left<  M^2\right>\right|_{O(G^0)} \ , \\
  \frac{\partial W_1}{\partial m}=\left.\left<  M\right>\right|_{O(G^2)},&& \quad 
  \frac{\partial W_1}{\partial \lambda}=\left.\left<  M^2\right>\right|_{O(G^2)} \ .
\nonumber
\end{eqnarray}
Inserting  the expectation values (\ref{eq:sol}) 
we get differential equations that determine  the 
dependence of the effective superpotential and the gravitational 
F-term on the tree level couplings. 

The differential equations at order $O(G^0)$ have been integrated in 
\cite{Brandhuber:2003va} and give the effective superpotential
\begin{eqnarray}
  \label{effpot}
  W_{eff}&=&N_cS\left(-\log{\frac{S}{\Lambda^3}} +1\right)+S \log{\frac{m}{\Lambda}} 
-S/2+S\log{\left(\frac{1}{2}+\frac{1}{2}\sqrt{1+\frac{8S\lambda}{m^2}}\right)} \\
& & -\frac{m^2}{8\lambda}+\frac{m^2}{8\lambda}\sqrt{1+\frac{8\lambda S}{m^2}} \ ,
\nonumber
\end{eqnarray}
where the integration constant has been fixed by matching to the Veneziano-Yankielowicz
potential. This means that $W_{eff}$ reproduces
the first line of (\ref{eq:vycorr}) in the limit
$\lambda\rightarrow 0$. 
Similarly, the equations at order $O(G^2)$ can be integrated 
to give $W_1$ 
\begin{eqnarray}
 W_1 & = & \frac{1}{3}\left(-N\log{m}-
\frac{2N-1}{2}\log{\left(1+\sqrt{1+\frac{8S\lambda}{m^2}}\right)}-
\frac{1}{4}\log{\left(1+\frac{8S\lambda}{m^2}\right)}\right)  \\
  &&+C(S,G^2) \ , \nonumber 
\end{eqnarray}
where the integration constant $C(S,G^2)$ is independent of the tree level couplings.  $C(S,G^2)$
is fixed by taking the limit $\lambda\rightarrow 0$ and requiring that
$W_1$ reproduces the second line of (\ref{eq:vycorr}).

With this matching the complete form of $W_1$ reads
\begin{eqnarray}
  \label{eq:effpot1}
  W_1&=&\frac{1}{3}\left(\frac{N_c^2-1}{2}\log\left(\frac{S}{\Lambda_0^3}\right)
-N_c\log{\frac{m}{\Lambda_0}}-
  \right.\\&&\left.-\frac{2N_c-1}{2}
\log{\left(\frac{1}{2}+\frac{1}{2}\sqrt{1+\frac{8S\lambda}{m^2}}\right)}-
 \frac{1}{4}\log{\left(1+\frac{8S\lambda}{m^2}\right)}\right) \ .
\nonumber
\end{eqnarray}

\bigskip\noindent{\it $SU(N_c)$  with $N_f$ Flavors}
\smallskip

To determine the gravitational couplings for the model with
$N_f$ fundamental flavors around the classical vacuum with
$N_f^-=N_f-N_f^+$ Higgsing quarks is only
slightly more complicated.
The details of this computation are given in
appendix  \ref{konrelnf}.

We recall first the result for $W_{eff}$ which has been computed in  \cite{Brandhuber:2003va}.
$W_{eff}$ reads
\begin{eqnarray}
\label{eq:efsNf}
W_{eff} & = & N_cS\left(-\log{\frac{S}{\hat{\Lambda}^3}}+1 \right) 
-N_f(\frac{S}{2}+\frac{m^2}{8\lambda})
+(N_f^+-N_f^-)\frac{m^2}{8\lambda}\sqrt{1+\frac{8\lambda}{m^2}S}+ \nonumber\\
& & +S\log\left[{\left(\frac{1}{2} +\frac{1}{2}
\sqrt{1+\frac{8\lambda}{m^2}S}\right)^{N_f^+}\left(\frac{1}{2} -\frac{1}{2}
\sqrt{1+\frac{8\lambda}{m^2}S}\right)^{N_f^-}}\right] ~.   
\end{eqnarray}
The scales involved  are the UV scale $\Lambda$, the scale 
$\hat\Lambda$ for the theory with $N_f$ massive quarks around 
$Q=0$, and the scale $\tilde\Lambda$ for the theory with $N_f^+$ 
massive quarks around $Q=0$ and $N_f^-$ Higgsing quarks,
\begin{eqnarray}
   \label{eq:mhq}
   \hat\Lambda^{3N_c}=\Lambda^{3N_c-N_f}m^{N_f}=
 \tilde\Lambda^{3(N_c-N_f^-)}(m^2/2\lambda)^{N_f^-} ~.
\end{eqnarray}
The order $O(G^2)$ contributions
to the relevant condensates in this vacuum are given by
\begin{equation}
\left. \left<tr M \right> \right|_{O(G^2)}=N_f^+M_1^+ + N_f^-M_1^- \ ,
\end{equation}
and
\begin{equation}
\left. \left<tr M^2 \right>\right|_{O(G^2)}=2 \left(N_f^+M_0^+M_1^++N_f^-M_0^-M_1^-\right)+\left(N_f^+\,^2 B^++N_f^-\,^2 B^-\right) \ ,
\end{equation}
where the $M_0,M_1,B$ with superscript $+\, (-)$ refer to the solutions of the Konishi relations (\ref{eq:compeq})
on the un-Higgsed (Higgsed) branch given in the appendix \footnote{See also \cite{Gripaios:2003gw} for a detailed analysis.}. Integrating with respect to $m$ and $\lambda$ and fixing the
integration constant in the way outlined above gives the gravitational F-term
\begin{eqnarray}
\label{fqs1}
W_1 & = & \frac{1}{6}\left[ \left(N_c-N_f^- \right)^2-1 \right] \log{\frac{S}{\Lambda_0^3}} 
-\frac{1}{6}\left[N_f^-N_f^- -2N_cN_f^- \right] \log{\frac{m}{-2\lambda\Lambda^2_0}} \nonumber \\
& & -\frac{1}{6}\left[N_f^-N_f^-+2N_cN_f^+\right] \log{\frac{m}{\Lambda_0}}-\frac{N_f^+\,^2+N_f^-\,^2}{12}\log{\left(1+\frac{8 \lambda S}{m^2}\right)}
\nonumber \\
& & +\frac{1}{6}\left(N_f^+-N_f^-\right)\left(N_f-2N_c \right)\log{\left(\frac{1}{2}+
\frac{1}{2}\sqrt{1+\frac{8\lambda S}{m^2}}\right)} \ .
\end{eqnarray}

 

\subsection{Parameter Space}
\label{branchesSQCD}

We now want to point out in what sense the perturbative part 
of $W_1$ gives us a consistency check on our approach to determine
the integration constant from the $U(1)_R$ anomaly. As an illustrative
example we will consider the classical vacuum with $N_f^-=0$ Higgsing quarks
and $N_f^+=N_f$ which corresponds to the vacuum with unbroken gauge group.
In the full quantum solution (\ref{fqs1}) this classical
vacuum is connected in the parameter space to the
classical vacuum with $N_f^-=N_f$ Higgsing quarks where the gauge group   
is maximally Higgsed from $SU(N_c)$ to $SU(N_c-N_f)$. 

By passing through the 
branch cut of the square root we can go from one 
semiclassical region to the other.
If the integration constant is indeed determined by the $U(1)_R$ anomaly
of the classically unbroken gauge group, which accounts for the strong coupling dynamics,
then changing the branch should provide the appropriate terms to adjust
the integration constant correctly. In other words for the vacuum with
unbroken gauge group we expect in the limit $\lambda \to 0$ the
gravitational F-term to be
\begin{equation}
W_1^{uh}=\frac{1}{3}\left(\frac{N_c^2-1}{2}
\log\left(\frac{S}{\Lambda_0^3}\right)-
N_fN_c\log{\frac{m}{\Lambda_0}} \right) \ .
\end{equation}
On the other hand for the maximally Higgsed vacuum we expect
\begin{eqnarray}
  W_1^h&=&\frac{1}{3}\left(\frac{(N_c-N_f)^2-1}{2}
\log\left(\frac{S}{\Lambda_0^3}\right)+\right.\\&&\left.
      +\left[N_f^2-2N_cN_f\right]\log\left(\frac{\Lambda_0}{m_h}\right)
      +\frac{N_f^2}{2}\log\left(\frac{\Lambda_0}{m}\right)\right) \ . \nonumber
\end{eqnarray}
The physical meaning of this potential is, 
that it comes from integrating out, first the massive $W$-bosons 
 at the Higgsing scale $m_h=\sqrt{-\frac{m}{2\lambda}}$
and then the remaining uncharged matter at the scale $m$.
As usual the dependence on the mass scales is determined
by integrating the Konishi relations but the important 
thing is that changing the branch of the square root in
(\ref{fqs1}) provides exactly the necessary terms to change
the coefficient of the $\log S$ term according to
\begin{equation}
\left(N_c^2-1\right)\log{S} \to  
\left(\left(N_c-N_f\right)^2-1\right)\log{S} \ ,
\end{equation}
which is precisely the expected behavior.
From the general solution (\ref{fqs1})
it is easy to see that the same is true for the general vacuum
with $N_f^- \ne 0$ Higgsing quarks.


\bigskip\noindent{\it Remarks On The Energy Scales}
\smallskip

We have seen that 
the functional dependence of the holomorphic couplings on $S$ 
can be fixed by anomaly matching. 
However, the scales $\Lambda_0$ appearing in the potential remain unfixed. 
More precisely, neither the Konishi nor the $U(1)_R$ anomaly 
restricts the scale in the $\log(S/\Lambda_0^3)$ to be 
the same as the scale in the $\log(m/\Lambda_0)$ term.

However, the complete solution fixes the two to be the same.
This can be seen as follows. 
The complete solution
allows to continue the gravitational coupling $W_1$ from the un-Higgsed vacuum 
to the Higgsed vacuum. This 
analytic continuation mixes the scale $\Lambda_0$ that saturates $S$ 
with the ones that saturate $m$ and $m_h$. Thus, these scales have to be the same.


\subsection{Comparison to the Vector Model}

The anomaly equation (\ref{KonishirelSQCD}) can also be derived from the zero 
dimensional vector model
\begin{eqnarray}
F({\hat N},N_f,g_s;m,\lambda)=-ln \int dQ d\tilde Q e^{-\frac{1}{g_s}W_{tree}(\tilde Q Q)} ~.
\end{eqnarray}
with $W_{tree}(\tilde Q Q)=m M+\lambda M^2$. The Ward identities for the variation $\delta Q=Q$  and 
$\delta Q=QM$  are
\begin{eqnarray}
 g_s{\hat N}&=&m \left< M\right>+\lambda \left< M^2\right>,\\  
 g_s{\hat N}\;\left< M\right>+g_s \left< M\right>&=&m \left< M^2\right>+\lambda \left< M^3\right>.
\end{eqnarray}
Making the identification $S-\frac{N_c}{3} G^2=g_s {\hat N}$ 
and $g_s=-\frac{1}{3} G^2$, one reproduces the Konishi anomaly equations (\ref{KonishirelSQCD}). 
Here ${\hat N}$ is the size of the vectors in the vector model.


\bigskip\noindent{\it Planar and Non-Planar Diagrams}
\smallskip


According to the proposal of Dijkgraaf and Vafa \cite{Dijkgraaf:2002dh}
the gravitational coupling $W_1(S)$ should be given by diagrams of genus one
in a corresponding matrix (vector) model. It was noted
in \cite{Ooguri:2003tt} that string theory arguments imply that there are also
planar contributions to  $W_1(S)$ which
vanish 
at the extrema of $W_{eff}(S)$. It has been shown 
in \cite{David:2003ke} that
in the adjoint model the planar contributions to $W_1$
can be generated from $W_{eff}$ by shifting $S$. This is not precisely the case 
in our model.

Consider the case $N_f=1$ with the gravitational F-term given by Eq. (\ref{eq:effpot1}).
We can identify the planar contributions to $W_{1}$ by shifting
$S \to S-\frac{1}{3}N_cG^2 \,$\footnote{The shift depends
on the matter representation, this is why in our case the shift differs
from the shift in the adjoint model \cite{David:2003ke}.} . 
This shift amounts to replacing one $S$ by $G^2$
in each of the planar diagrams and the factor of $N_c$ accounts for the
index loop without an $S$.
The appropriate combinatorial factor which is induced by this procedure 
(for each diagram we now have to choose an additional index loop to be free of $S$)
is taken care of by the derivative with respect to  $S$.
In this sense we have
\begin{equation}
W_{1}^{planar}=\frac{\partial W_{eff}}{\partial S} \delta S \ ,
\end{equation}
with $\delta S=-\frac{N_c}{3}G^2$.

Consider the perturbative part of $W_{eff}$  (\ref{effpot}).
Shifting $S$ gives:
\begin{equation}
\label{W1again}
-\frac{N_c}{3} \log \frac{m}{\Lambda} - \frac{N_c}{3} \log \left(\frac{1}{2}
+\frac{1}{2}\sqrt{1+\frac{8\lambda S}{m^2}} \right) \ . 
\end{equation}
We can identify this part in $W_1$ with the identification of the scales
$\Lambda_0 = \Lambda$.
The perturbative non-planar part of $W_1$ is then
\begin{equation}
\label{W1again1}
-\frac{1}{12}\log \left(1+\frac{8\lambda S}{m^2}\right)
+\frac{1}{6}\log 
\left(\frac{1}{2}+\frac{1}{2} \sqrt{1+\frac{8\lambda S}{m^2}}
\right) \ .
\end{equation}
However, it is easy to see that the shift of $S$ in the non-perturbative part of $W_{eff}$
does not reproduce the non-perturbative planar part  of $W_1$
that goes like $N_c^2$.

\section{A Chiral $SU(6)$ Model}

We consider in this section 
the chiral $SU(6)$ model
analysed in \cite{Brandhuber:2003va}.
The matter content of this model consists of
two antifundamental flavors $\bar{Q}^I_i$ and one flavor $X^{ij}$ in
the antisymmetric representation. 
Capital $I,J=1,2$ denote flavor indices and $i,j$ denote
color indices.
The
relevant gauge invariant operators are
\begin{equation}
T=\varepsilon_{IJ}\bar{Q}^I_i\bar{Q}^J_jX^{ij} \ ,
\end{equation}
and
\begin{equation}
U=Pf X= X^{i_1 j_1}X^{i_2 j_2} X^{i_3 j_3}\varepsilon_{i_1 j_1i_2 j_2i_3 j_3} \ .
\end{equation}
We will consider the tree level superpotential
\begin{equation}
W_{tree}=hT+gU+\lambda TU \ ,
\end{equation}
where, as explained in detail in  \cite{Brandhuber:2003va}, the term $\lambda TU$
will provide us with a region of the quantum parameter space where, despite the absence
of a massive vacuum,  we can 
reliably determine the integration constant by matching the $U(1)_R$ anomaly. 

Again, we have to take the effect of the 
supergravity background into account. This leads to the following
general expansion rules for the chiral correlators
\begin{eqnarray}
\label{Gexp1}
\left<A \right> &  \equiv & A_0 +G^2 A_1 \ , \\ 
\left<AB \right> &  \equiv & \left<A \right> \left<B \right>+
G^2\left<AB \right>_c\ . \nonumber 
\end{eqnarray}
Concerning the three point correlators we will argue that the connected part 
vanishes in the chiral ring 
as in the cases of SYM with non-chiral matter  
in the fundamental representation and the adjoint model \cite{David:2003ke}.
The diagrammatic proof seems to be more intricate in this case and we will
instead use the Konishi equations to show perturbatively in $\lambda$ (or $S$) 
that the connected part $\left<ABC \right>_c$
of any three point correlator vanishes. 
We can then find a closed system of equations 
which allows to solve for  $T_{0,1},U_{0,1}$ and $\left<TU \right>_c$ 
in order to determine the dependence of the gravitational F-terms on the tree level couplings.


\subsection{Connected $n$-Point Correlators}

Our strategy will be the following. We will start by
deriving the most general Konishi relations based on the variations
\begin{equation}
\label{varis}
\delta A = A
T^nU^m \ ,
\end{equation}
where $A$ stands for either $\bar{Q}$ or $X$.
We will assume that the connected parts of all the
$n$-point functions for $n \ge 3$ have nonvanishing contributions
at order $O(G^2)$. 
It is then easy to verify that for $n+m > 1$ the Konishi
relations can generate a closed system of
equations for the correlators. However, these
equations turn out to be linearly dependent and we will resort
to the following perturbative argument.
First, we are going to show by induction that
for $\lambda \to 0$ the connected parts of the $n$-point correlators on the un-Higgsed branch 
vanish for $n > 2$. On dimensional grounds we then have 
\begin{equation}
\label{lambdarel1}
\left<T^n U^m\right>_c=\lambda S^{n+m} f\left(\lambda S \right) \ ,
\end{equation}
where in the limit $\lambda \to 0$ the function $f$ has to go 
with some positive (or zero) power of $\lambda$.
We then turn on $\lambda$. Due to the linear dependence
in this case we can only solve in terms of one unknown, say, 
$T_1$ defined in the expansions (\ref{Gexp1}). 
By requiring that the correlator $\left<T^a U^b \right>$
show the behavior of (\ref{lambdarel1}) we can solve for
$T_1$ up to $(a+b-1)$'th order in $\lambda$.
We find perfect agreement with the expansion of the complete solution
which we obtain when we assume that the connected three point correlators
$\left<ABC \right>_c$ vanish.

\bigskip\noindent{\it The Model With $\lambda \to 0$}
\smallskip

If we set $\lambda \to 0$ we can solve explicitly for
$T_{0,1},U_{0,1}$ and $\left<TU\right>_c,\left<TT\right>_c,
\left<UU\right>_c$. The solution is given by
\begin{eqnarray}
\label{sol1}
T_0=\frac{S}{h} \, & & T_1=-\frac{2}{h} \ , \nonumber \\
U_0=\frac{S}{g} \, & & U_1=-\frac{1}{g} \ , \nonumber \\
\left<TT\right>_c = -\frac{S}{3h^2} \ , &  & 
\left<TU\right>_c = 0 \ , \nonumber \\
\left<UU\right>_c = -\frac{S}{3g^2} \ . 
\end{eqnarray}
The variations (\ref{varis})
lead to
\begin{equation}
\label{lnull1}
h\left<T^{n+1}U^m\right> = S \left<T^n U^m \right>-\frac{G^2}{3}\left(n+6 \right)
\left<T^nU^m\right> \ ,
\end{equation}
for $\delta \bar{Q}_i^1 = \bar{Q}_i^1T^n U^m $
and 
\begin{equation}
\label{lnull2}
h\left<T^{n+1}U^m\right>+3g\left<T^{n}U^{m+1}\right> = 4S \left<T^n U^m \right>
-\frac{G^2}{3}\left(n+3m+15 \right)
\left<T^nU^m\right> \ ,
\end{equation} 
for $\delta X^{ab} = X^{ab}T^n U^m $.
Using the above equations we can verify that the connected correlators $\left<T^nU^m \right>_c$
vanish for $n+m>2$. First, plugging the solutions (\ref{sol1}) into the equations
(\ref{lnull1}) and (\ref{lnull2}) and assuming that all the connected
correlators up to $\left<T^nU^m \right>_c$ for $n+m$ fixed vanish we find that
the correlators $\left<T^{n+1}U^m \right>_c$ and
$\left<T^nU^{m+1} \right>_c$ also vanish. 
Verifying by explicit calculation that all the connected three point correlators
vanish finally allows us to conclude that also all higher connected correlators
vanish. For finite $\lambda$ dimensional analysis
then restricts the correlators to be of the form given in
(\ref{lambdarel1}).

\bigskip\noindent{\it The Model With Finite $\lambda$}
\smallskip

For a finite value of $\lambda$ the Konishi relations (\ref{lnull1}),(\ref{lnull2})
get modified to 
\begin{eqnarray}
\label{lfinite1}
h\left<T^{n+1}U^m\right> +\lambda \left<T^{n+1}U^{m+1}\right>& = & 
S \left<T^n U^m \right>-\frac{G^2}{3}\left(n+6 \right)\left<T^n U^m \right>  \ , \nonumber \\
h\left<T^{n+1}U^m\right>+3g\left<T^{n}U^{m+1}\right> +4\lambda \left<T^{n+1} U^{m+1} \right>& = & 
4S \left<T^n U^m \right> \nonumber\\
-\frac{G^2}{3}\left(n+3m+15 \right)
\left<T^nU^m\right> \ . 
\end{eqnarray}
If we want to include all the $k$-point functions we have to consider all
equations with $n+m+1 \le k$. The $\lambda$ term in the Konishi relations
also introduces some $k+1$ point functions into the game. The solution
of this set of equations will be given in terms of one of the variables
which we take to be $T_1$. Expanding $T_1$ in a series of $\lambda$
and imposing on the $k+1$-point functions that they go like
$\lambda S^{k+1}$ for small $\lambda$ determines $T_1$ up to order
$k$ in $\lambda$. We have checked explicitly
up to order $O(\lambda^8)$. The result we get for $T_1$ is given by
\begin{eqnarray}
\label{t1}
T_1 & = & -\frac{2}{h}+\frac{3S}{gh^2}\lambda -\frac{28S^2}{3g^2h^3}\lambda^2
+\frac{32S^3}{g^3h^4}\lambda^3 -\frac{344S^4}{3g^4h^5}\lambda^4 \nonumber \\
 & & +\frac{1264S^5}{3g^5h^6}\lambda^5-\frac{4720S^6}{3g^6h^7}\lambda^6
+\frac{17824S^7}{3g^7h^8}\lambda^7-\frac{67864S^8}{3g^8h^9}\lambda^8+O(\lambda^9)
\ .
\label{pseries}
\end{eqnarray}
Below we will see that this is in perfect agreement with the complete solution
when we take the connected three point correlators to vanish. We expect this 
argument to be applicable to any order in $\lambda$ provided one solves
(\ref{lfinite1}) up to high enough order in $n+m$.

\bigskip\noindent{\it The Complete Solution}
\smallskip

If we assume the connected three point correlators to vanish we can solve
for $T_{0,1},U_{0,1}$ and $\left<TU \right>_c$ in order to
determine the dependence of the gravitational F-terms on the tree level couplings.
The details of this calculation including the appropriate anomalous
transformations and the corresponding Konishi equations are given in appendix \ref{chiralapp}.
We obtain the following results
\begin{eqnarray}
\label{results1}
T_0 & = & -\frac{g}{2 \lambda} \left(1 \mp \sqrt{1+\frac{4 \lambda}{gh}S} \right) \ , \nonumber \\
U_0 & = & -\frac{h}{2 \lambda} \left(1 \mp \sqrt{1+\frac{4 \lambda}{gh}S} \right) \ , \nonumber \\
T_1 & = & -\frac{2}{3h} \mp \frac{4}{3h\sqrt{1+\frac{4 \lambda}{gh}S}}
+\frac{\lambda S}{3gh^2\left(1+\frac{4 \lambda}{gh}S\right)} \ , \nonumber \\
U_1 & = & \frac{1}{3g} \mp \frac{4}{3g\sqrt{1+\frac{4 \lambda}{gh}S}}
+\frac{\lambda S}{3g^2h\left(1+\frac{4 \lambda}{gh}S\right)} \ , \nonumber  \\
<TU>_c & = & \frac{1}{6\lambda} \left(\pm \frac{gh+2 \lambda S}{gh \sqrt{1+\frac{4 \lambda}{gh}S}}-1 \right) \ , \nonumber \\
<TT>_c & = & \mp \frac{S}{3h^2 \sqrt{1+\frac{4 \lambda}{gh}S}} \ , \nonumber \\
<UU>_c & = & \mp \frac{S}{3g^2 \sqrt{1+\frac{4 \lambda}{gh}S}} \ ,
\end{eqnarray}
where the (upper) lower sign is valid on the (un)Higgsed branch.
Expanding $T_1$ on the un-Higgsed branch with respect to  $\lambda$
one finds that it agrees with what we have found in
(\ref{t1}). We consider this a convincing argument that the assumption 
about the vanishing connected three point correlators is correct. 

It was shown in \cite{Brandhuber:2003va} that on the Higgsed branch the theory
classically reduces to a pure $Sp(2)$ theory. This fact made it possible
to determine the full effective superpotential $W_{eff}$
by matching to the Veneziano-Yankielowicz potential,
\begin{eqnarray}
W_{eff} & =& -5S\log{\frac{S}{\Lambda_6^3}} + 4S -2S \log{2}+ \nonumber \\
&& + S\log{gh}+\frac{gh}{2\lambda}\left(-1\pm
\sqrt{1+\frac{4\lambda}{gh}S}\right)+ \nonumber \\
&& + 2S\log\left(\frac{1}{2}\pm\frac{1}{2}\sqrt{1+\frac{4\lambda}{gh}S}\right) \ .
\end{eqnarray}
Similarly, we will determine
here the integration constant for the gravitational F-terms by matching
the perturbative solution on the Higgsed branch to the gravitational part of the $U(1)_R$
anomaly 
for pure $Sp(2)$.
Integrating the order $O(G^2)$ parts of $\left<T \right>,\left<U \right>$ and
$\left<TU \right>$, as indicated in (\ref{Gexp1}), with respect to $h,g$ and $\lambda$ 
yields
\begin{eqnarray}
W_1^h & = & -\log{gh^2}-\frac{8}{3}\log{\frac{gh}{-\lambda \Lambda_0^3}}
-\frac{1}{12}\log{\left(1+\frac{4 \lambda}{gh}S\right)} \nonumber \\
& & +\frac{8}{3}\log{\left(1 +\sqrt{1+\frac{4 \lambda}{gh}S}\right)}+C(S) \ .
\end{eqnarray}
To determine $C(S)$ we require that in the limit $S \to 0$ the leading
$\log S$ term of $W_1(S)$ reproduce the gravitational $U(1)_R$ anomaly for pure $Sp(2)$.
In the conventions we have used throughout the paper this is given by
\begin{equation}
\frac{1}{6}\mathrm{dim}\left({\mathbf{Adj}}_{Sp(2)} \right) \log S = \frac{5}{3} \log S \ .
\end{equation}
The full solution on the Higgsed branch then reads
\begin{eqnarray}
W_1^h & = & \frac{5}{3} \log{\frac{S}{\Lambda_0^3}} -\log{gh^2}-\frac{8}{3}\log{\frac{gh}{-\lambda \Lambda_0^3}}
-\frac{1}{12}\log{\left(1+\frac{4 \lambda}{gh}S\right)} \nonumber \\
& & +\frac{8}{3}\log{\left(\frac{1}{2} +\frac{1}{2}\sqrt{1+\frac{4 \lambda}{gh}S}\right)} \ .
\end{eqnarray} 
As usual, the un-Higgsed branch of the parameter space can
be reached by passing through the branch cut of the square root
in $W_1$, which leads us to
\begin{eqnarray}
W_1^{uh} & = & \frac{13}{3} \log{\frac{S}{\Lambda_0^3}} -\log{gh^2}
-\frac{1}{12}\log{\left(1+\frac{4 \lambda}{gh}S\right)} \nonumber \\
& & -\frac{8}{3}\log{\left(\frac{1}{2} +\frac{1}{2}\sqrt{1+\frac{4 \lambda}{gh}S}\right)} \ .
\end{eqnarray} 
Expanding the derivative with respect to $h$ of $W_1^{uh}$ as a power series
in the coupling $\lambda$ reproduces (\ref{pseries}).

\section*{Acknowledgements}

We would like to thank M. Mari\~no and T. Taylor for valuable discussions.
We would also like to thank the TH-Division in CERN for hospitality. 
This research is supported by the US-Israel Binational Science
Foundation and the TMR European Research Network.

\newpage
\appendix

\section{Konishi relations SYM with $N_f\neq 1$}
\label{konrelnf}
The Konishi equations relevant for the perturbation under consideration are
\begin{eqnarray}
\label{konishirelSQCD}
  \delta Q_i^f=Q_i^{f'}:&&(S-\frac{N_c}{3}G^2)
\delta_f^{f'} =m \left< M_f^{f'}\right>+2\lambda \left< (M^2)_f^{f'}\right>,\\
  \delta Q_i^f=Q_i^{f'}M_h^{h'}:&&(S-\frac{N_c}{3}G^2)
\delta_f^{f'}\left< M_h^{h'}\right>=m \left< M_f^{f'}M_h^{h'}\right>+2\lambda 
\left< (M^2)_f^{f'}M_h^{h'}\right> \nonumber\\       
&&\quad\quad\quad+    \frac{1}{3} \delta^{h'}_{f}\left< M_h^{f'}\right> G^2,
\end{eqnarray}
We study these equations for the case where the flavor symmetry is not broken. With the conventions
\begin{eqnarray}
\left< M_f^{f'}\right>&=&\delta_f^{f'} M,\\
\left< M_f^{f'}M_h^{h'}\right>&=&\left< M_f^{f'}M_h^{h'}\right>_c+\left< M_f^{f'}\right>\left< M_h^{h'}\right>,\\
\left< M_f^{f'}M_h^{h'}\right>_c&=&\delta_f^{f'}\delta_h^{h'} A+\delta_f^{h'}\delta_h^{f'} B,\\
\left< M_f^{f'}M_h^{h'}M_g^{g'}\right>&=&\left< M_f^{f'}M_h^{h'}\right>_c\left< M_g^{g'}\right>+
\left< M_f^{f'}M_g^{g'}\right>_c\left< M_h^{h'}\right> \\
& & +\left< M_h^{h'}M_g^{g'}\right>_c\left< M_f^{f'}
\right>+\left< M_f^{f'}\right>\left< M_h^{h'}\right>\left< M_g^{g'}\right>, \nonumber \\
\left< (M^2)_f^{f'}M_h^{h'}\right>&=&\delta_f^{f'}\delta_h^{h'}(3AM+N_fBM+M^3)+\delta_f^{h'}
\delta_h^{f'}2BM \ .
\end{eqnarray}
The Konishi equations then are
\begin{eqnarray}
  \label{eq:konflav}
  S&=&mM+2\lambda(M^2+A+N_fB)+\frac{N}{3}G^2,\\
  SM&=&m(A+M^2)+2\lambda\left[3AM+N_fBM+M^3\right]+\frac{N}{3}M\;G^2,\\
  0&=&mB+4\lambda BM+\frac{1}{3}MG^2 \ ,
\end{eqnarray}
where
\begin{eqnarray}
  M&=&M_0+M_1G^2,\\
  A&=&A_2G^4,\\
  B&=&B_1G^2 \ .
\end{eqnarray}
In components 
\begin{eqnarray}
  \label{eq:compeq}
  S&=&mM_0+2\lambda M_0^2,\\
  0&=&mM_1+2\lambda(2M_0M_1+N_fB_1)+\frac{N}{3},\\
  0&=&mB_1+4\lambda B_1M_0+\frac{M_0}{3} \ ,
\end{eqnarray}
and the redundant equations
\begin{eqnarray}
  SM_0&=&mM_0^2+2\lambda M_0^3,\\
  SM_1&=&2mM_0M_1+2\lambda(N_fB_1M_0+3M_0^2M_1)+\frac{M_0}{3} \ .
\end{eqnarray}
The solution is given by
\begin{eqnarray}
M_0^{\pm} & = & \frac{-m \pm \sqrt{m^2+8 \lambda S}}{4 \lambda} \ , \\
M_1^{\pm} & = & \mp \frac{N_c}{3\sqrt{m^2+8\lambda S}}+
2\lambda N_f\frac{-m \pm \sqrt{m^2+8 \lambda S}}{12\lambda  
\left(m^2+8 \lambda S \right)} \ , \nonumber \\
B_1^{\pm} & = & \mp\frac{-m \pm \sqrt{m^2+8 \lambda S}}
{12 \lambda \sqrt{m^2+8 \lambda S}} \ .
\end{eqnarray}

\section{Details of the Chiral Model}
\label{chiralapp}

The transformation
\begin{equation}
\delta {\bar Q}^1_i= {\bar Q}^1_i 
\end{equation}
gives the chiral ring relation
\begin{equation}
hT+\lambda TU = S-\frac{G^2}{3}N \ ,
\end{equation}
for $N=6$. Here and in the following Konishi relations we will use $N$ and replace it only 
later by the actual value.
If we expand the one point functions in $G^2$ and 
take into account the connected correlators we find 
\begin{equation}
O(G^0) \; : \; \;hT_0 + \lambda T_0 U_0=S \ ,
\end{equation}
and
\begin{equation}
O(G^2) \; : \; \;hT_1 + \lambda \left(T_1 U_0+ T_0U_1 +\left<TU\right>_c \right)=-\frac{N}{3} \ .
\end{equation}
The chiral ring relation induced by the transformation
\begin{equation}
\delta X^{mn}= X^{mn}
\end{equation}
is given by
\begin{equation}
hT +3gU+4 \lambda TU=\left(N-2\right)S-\frac{G^2}{3}\frac{N\left(N-1\right)}{2} \ .
\end{equation}
At order $O(G^0)$ and $O(G^2)$ we find respectively
\begin{equation}
O(G^0) \; : \; \;hT_0 +3gU_0+4 \lambda T_0U_0=\left(N-2\right)S \ ,
\end{equation}
\begin{equation}
O(G^2) \; : \; \;hT_1 +3gU_1+4 \lambda \left(T_1U_0+T_0U_1+\left<TU\right>_c\right)=-\frac{N\left(N-1\right)}{6} \ .
\end{equation}
From the two equations at order $O(G^2)$ it follows that $U_1$ and $T_1$
satisfy
\begin{equation}
gU_1-hT_1=1 \ .
\end{equation}
(for $N=6$).
The variation 
\begin{equation}
\delta {\bar Q}^1_i= {\bar Q}^1_i T 
\end{equation}
gives rise to
\begin{equation}
hT^2+\lambda T^2U=ST-\frac{G^2}{3}\left(N+1\right)T \ ,
\end{equation}
which gives the following relevant equation at order $O(G^2)$
\begin{eqnarray}
O(G^2):& & 2hT_0T_1+h\left<TT\right>_c \nonumber \\
 & & +\lambda \left(T_0^2U_1+T_0T_1U_0+T_1T_0U_0 +U_0\left<TT\right>_c+2T_0\left<TU\right>_c\right)
\nonumber \\ 
 & & +\lambda \left<TTU\right>_c\,=ST_1-\left(N+1\right)\frac{T_0}{3} \ .
\end{eqnarray}
Similarly, for the variation 
\begin{equation}
\delta {\bar Q}^1_i= {\bar Q}^1_i U 
\end{equation}
we find 
\begin{eqnarray}
O(G^2):& & hT_0U_1+hT_1U_0+h\left<TU\right>_c \nonumber \\
& & +\lambda \left(2T_0U_0U_1+T_1U_0U_0 +T_0\left<UU\right>_c+2U_0\left<TU\right>_c\right)
\nonumber \\ 
& & +\lambda \left<TUU\right>_c\,=SU_1-\frac{N}{3}U_0 \ .
\end{eqnarray} 
Finally, 
\begin{equation}
\delta X^{mn} = X^{mn}T
\end{equation} 
leads to
\begin{eqnarray}
O(G^2):& & 2hT_0T_1+h\left<TT\right>_c+3gT_0U_1+3gU_0T_1 \nonumber \\
 & & + 3g\left<TU\right>_c+4\lambda \left(T_0T_0U_1+2T_0T_1U_0\right)
\nonumber \\ 
& & +8\lambda T_0\left<TU\right>_c+4\lambda U_0\left<TT\right>_c+4\lambda \left<TTU\right>_c\,= \nonumber \\
& & \left(N-2\right)ST_1-\frac{T_0}{6}\left[N\left(N-1\right)+2\right] \nonumber \ .
\end{eqnarray} 
We derive another Konishi relation which comes from the variation
\begin{equation}
\delta X^{mn} = X^{mn}U \ .
\end{equation}
The corresponding relation in the chiral ring reads
\begin{equation}
hTU+3gUU+4 \lambda TUU = \left(N-2\right)SU-G^2\frac{U}{6}\left[N\left(N-1\right)+6\right] \ ,
\end{equation}
which at order $O(G^2)$ gives
\begin{eqnarray}
O(G^2):& & hT_0U_1+hT_1U_0+h\left<TU\right>_c+6gU_0U_1 \nonumber \\
& & + 3g\left<UU\right>_c+4\lambda \left(T_1U_0U_0+2T_0U_0U_1\right)
\nonumber \\ 
& & +4\lambda T_0\left<UU\right>_c+8\lambda U_0\left<TU\right>_c + 4\lambda\left<TUU\right>_c \nonumber \\
& & =\left(N-2\right)SU_1-\frac{U_0}{6}\left[N\left(N-1\right)+6\right] \ .
\end{eqnarray}



\end{document}